\documentclass[journal=cmatex,layout=draft]{achemso}
\setkeys{acs}{
    articletitle = false,
}
\pdfoutput=1 
\usepackage{units}
\usepackage{graphicx}
\usepackage[colorlinks,hyperindex]{hyperref}
\usepackage[hyperindex]{hyperref}
\hypersetup{colorlinks,citecolor=blue,linkcolor=blue,urlcolor=blue}
\usepackage{amsmath}
\usepackage[version=4]{mhchem}
\usepackage[caption=false]{subfig}
\usepackage{lipsum}
\usepackage{siunitx}

\author{Dorothea Scheunemann}
\email{dorothea.scheunemann@uol.de}
\affiliation{Complex Materials and Devices, Department of Physics, Chemistry and Biology (IFM), Linköping University, 58183 Linköping, Sweden}
\author{Martijn Kemerink}
\affiliation{Complex Materials and Devices, Department of Physics, Chemistry and Biology (IFM), Linköping University, 58183 Linköping, Sweden}
\affiliation{Centre for Advanced Materials, Heidelberg University, Im Neuenheimer Feld 225, 69120 Heidelberg, Germany}

\title{Non-Wiedemann-Franz Behavior of the Thermal Conductivity of Organic Semiconductors}

\begin{document}

\begin{abstract}
Organic semiconductors have attracted increasing interest as thermoelectric converters in recent years due to their intrinsically low thermal conductivity compared to inorganic materials. This boom has led to encouraging practical results, in which the thermal conductivity has predominantly been treated as an empirical number. However, in an optimized thermoelectric material, the electronic component can dominate the thermal conductivity in which case the figure of merit $ZT$ becomes a function of thermopower and Lorentz factor only. Hence design of effective organic thermoelectric materials requires understanding the Lorenz number. Here, analytical modeling and kinetic Monte Carlo simulations are combined to study the effect of energetic disorder and length scales on the correlation of electrical and thermal conductivity in organic semiconductor thermoelectrics. We show that a Lorenz factor up to a factor $\sim 5$ below the Sommerfeld value can be obtained for weakly disordered systems, in contrast with what has been observed for materials with band transport. Although the electronic contribution dominates the thermal conductivity within the application-relevant parameter space, reaching $ZT>1$  would require to minimize both the energetic disorder but also the lattice thermal conductivity to values below $\kappa_\text{lat}<\SI{0.2}{W/mK}$. 
\end{abstract}

\maketitle

\section{Introduction}
\label{sec:intro}
Semiconducting organic materials have attracted increasing interest as thermoelectric (TE) converters in the recent years due to their potentially low material and fabrication costs \cite{LeBlanc2014} and non-toxicity \cite{Sun2019,Lu2016_2}. Another key benefit of organic semiconductors compared to typical inorganic materials is their intrinsically low thermal conductivity that is due to structural disorder and weak intermolecular coupling. The latter can remarkably increase the thermoelectric conversion efficiency, governed by the dimensionless figure of merit 
\begin{equation}
ZT=\frac{S^2\sigma T}{\kappa},
\label{eq:ZT}
\end{equation}
where $S$ is the Seebeck coefficient, $\sigma$ is the electrical conductivity, $T$ is the absolute temperature, and 
\begin{equation}
\kappa=\kappa_\text{el}+\kappa_\text{lat}
\label{eq:kappa}
\end{equation}
is the total thermal conductivity, composed of electronic ($\kappa_\text{el}$) and lattice ($\kappa_\text{lat}$) contributions.
For inorganic materials, it is well known that the parameters $S$, $\sigma$ and $\kappa$ are coupled, so that optimizing one parameter tends to compromise the others. This makes the optimization of the TE efficiency an extremely challenging task, requiring strategies such as engineering of the electronic band structure and lattice thermal conductivity.

Such correlations are less established for organic semiconductors. In the recent years, lots of effort in the field of organic TEs went into studying the relation between $\sigma$ and $S$, both experimentally and theoretically \cite{Wang2012,Lu2016_2,Abdalla2017,Zuo2018}. However, little is known about the correlation between electrical and thermal conductivity in these materials. In inorganic materials, the electronic contribution of the thermal conductivity is related to the electrical conductivity by the Wiedemann-Franz law, $\kappa_\text{el}/\sigma=LT$, with a Lorenz number $L$ that is typically close to the Sommerfeld value $L_0 = \frac{\pi^2}{3}\left(\frac{k_\text{B}}{q}\right)^2$ \cite{Sommerfeld1927}.
In contrast, for organic materials there are indications that the Wiedemann-Franz law does not hold, and a large range of Lorenz numbers has been reported. While some experimental studies suggest $L$ to be equal to the Sommerfeld value for a free electron gas \cite{Salamon1975,Liu2015}, others show large deviations from this value in both directions \cite{Yoon1991,Kim2013,Weathers2015}. This experimental divergence is accompanied by a lack of theoretical understanding. Lu \textit{et al}. \cite{Lu2016} as well as Upadhyaya \textit{et al}. \cite{Upadhyaya2019} reported large deviations from the Sommerfeld value, which was mainly attributed to the effect of disorder but the results are contradictory. The low attention paid to the electronic contribution of the thermal conductivity and  Lorenz factor is surprising. First, in the application relevant regime the electronic contribution can dominate the thermal conductivity. Second, in the limit when electronic thermal conductivity dominates $\kappa_\text{lat}\ll\kappa_\text{el}$, the maximum $ZT$ for any material is given by $ZT = S^2/L$. Thus, reducing the Lorenz number will be crucial for maximum $ZT$.

Here, we combine kinetic Monte Carlo (kMC) simulations and analytical modeling to study the electronic thermal conductivity $\kappa_\text{el}$ of disordered organic semiconductors. All simulations are based on a hopping formalism that takes into account the specific shape of the density of states (DOS). In particular, we also investigate a Gaussian DOS modified by Coulomb trapping of mobile charge carriers by ionized dopants. We show that the presence of energetic disorder as well as the magnitude of the localization length can lead to significant derivations from the Wiedemann-Franz law, corresponding to effective Lorenz numbers that can either be smaller or larger than $L_0$.

\section{Theoretical Framework}
\label{sec:model}
The macroscopic observables $\sigma$, $S$ and $\kappa_\text{el}$ can be determined from\cite{Ziman}:
\begin{equation}
\sigma = \int_{-\infty}^{\infty} dE\, \sigma(E)\left(-\frac{\partial f_\text{FD}}{\partial E}\right) = \int_{-\infty}^{\infty} dE\, \sigma'(E),
\label{eq:sigma}
\end{equation}
\begin{equation}
S=\frac{1}{qT}\int dE\, (E-E_F) \frac{\sigma'(E)}{\sigma},
\label{eq:S}
\end{equation}
\begin{equation}
\kappa_0=\frac{1}{q^2T}\int dE\, (E-E_F)^2 \sigma'(E)
\label{eq:kappa0}
\end{equation}
with the electronic thermal conductivity defined as 
\begin{equation}
\kappa_\text{el} = \kappa_0-S^2\sigma T.
\label{eq:kappael}
\end{equation}
Here $f_\text{FD}$ is the Fermi-Dirac distribution function, $E_F$ the Fermi level, $\sigma'(E)$ the energy-dependent differential conductivity and $\kappa_0$ the electronic thermal conductivity when the electrochemical potential gradient inside the sample is zero. 
Eq.~(\ref{eq:sigma})-(\ref{eq:kappa0}) can be derived from the Boltzmann equation in the relaxation time approximation. Although it is up to now uncertain whether this approximation after the transition from momentum to energy space is also valid for disordered organic semiconductor materials, H. Fritzsche \cite{Fritzsche1971} showed that Eq.~(\ref{eq:sigma}) and (\ref{eq:S}) are generally applicable, independent of a specific conduction process. Furthermore, it was shown by Gao \textit{et al.} that Seebeck coefficients calculated using Eq.~(\ref{eq:S}) are in good agreement with experimental results on crystalline polymers\cite{Gao2005}. In analogy, Eq.~(\ref{eq:kappa0}) is here used as an ansatz and its applicability is discussed below.

\subsection{Numerical Model}
The kMC model has been extensively described  before\cite{Abdalla2015,Zuo2019}. In brief, the kMC simulations account for variable-range hopping on a random lattice with a mean intersite distance of $a_\text{NN}=N_\text{0}^{-1/3}=\SI{1.8}{nm}$ with $N_\text{0}$ the total site density. Site energies are distributed according to a Gaussian DOS with varying degree of disorder. To account for doping, Coulomb interactions with all charged particles are included. 

The thermal conductivity $\kappa_\text{el}$ was calculated numerically exact using the definitions in Eq.~(\ref{eq:sigma})--(\ref{eq:kappael}). 
Using $\sigma'(E)=j(E)/F$, where $F$ is the constant electric field and $j(E)$ the current density at an energy $E$, we arrive at:
\begin{multline}
\kappa_0=\frac{1}{q^2T}\left(\int dE\,E^2\, \frac{j(E)}{F} - 2 E_F \int dE\,E\, \frac{j(E)}{F}\right.\\
\left. + E_F^2 \int dE\, \frac{j(E)}{F} \right).
\label{eq:kappa0_2}
\end{multline}
To handle integrals of the form $\int dE\,E^x\,j(E)$ with $x=[0-2]$, we use the fact that under steady-state conditions all (differential) currents are constant and integrate over time. The resulting double integrals reflect the total amount of charge that passes through a unit cross section in a given time $t$, weighted by $E^x$. Numerically, they can be evaluated as sums over all hopping events $i$ in the time $t$,
\begin{equation}
\int_{0}^{t}\int E^x j(E)\,dE\,dt = \frac{1}{AL} \sum_i E_i^x\,q \Delta z_i,
\label{eq:sum}
\end{equation}
where $A$ and $L$ are the cross section and length of the simulation box, respectively,
and $\Delta z_i$ the displacement of the $i$th hopping event in the direction of the electric field applied along $z$. The energy of a hop between sites i and j is calculated as $(E_\text{i}+E_\text{j})/2$ The thermopower $S$ is determined analogously, further details can be found in Ref.~\cite{Zuo2019}.

\subsection{Analytical Model}
Our analytical model extends the work of Schmechel \cite{Schmechel2003} and Ihnatsenka \textit{et al}.,\cite{Ihnatsenka2015} that expresses the Seebeck coefficient for a hopping system in a differential conductivity $\sigma'(E)$ to also describe the heat conductivity of the hole or electron gas. For completeness, the key expressions are given in the following. To calculate the total conductivity $\sigma$, thermopower $S$ and the electronic contribution to the thermal conductivity $\kappa_\text{el}$, the resulting differential conductivity is inserted in expressions for the transport coefficients stemming from the Boltzmann transport equation (see Eq.~(\ref{eq:sigma})--(\ref{eq:kappael})).

The differential escape rate distribution $\nu_\text{esc}'$ from a state with energy $E_0$ via a state with $E_0+W$ is given by the product of the thermal activation rate from the initial state at energy $E_0$ to an intermediate state at energy $E_0 +W$ and the tunneling rate from the intermediate state to a final state below $E_0$ according to the Miller-Abrahams-type expression:
\begin{equation}
\nu_\text{esc}'(E_0,W)=\frac{\nu_0}{k_\text{B}T}\exp\left(-\frac{2\bar{R}(E_0+W)}{\alpha}\right)\exp\left(-\frac{W}{k_\text{B}T}\right).
\label{eq:vesc1}
\end{equation}
Here, $\nu_0$ is the carrier attempt-to-hop frequency, $k_\text{B}$ is the Boltzmann constant, $T$ is the temperature and $\alpha$ describes the decay length of the localized wavefunction. It was shown in Ref.~\cite{Schmechel2003} that Eq.~(\ref{eq:vesc1}) is equivalent to the common Miller-Abrahams expression if the total transfer rate from an initial state with energy $E_0$ to a specific target state at energy $E_\text{target}$, separated by a distance $R$ is calculated. The parameter $\bar{R}(E)$ is the mean tunneling distance for an electron at energy $E$ to reach a target state at lower energy:
\begin{equation}
\bar{R}(E)=\left(\frac{4\pi}{3B}\int_{-\infty}^{\text{E}}d\epsilon\, g(\epsilon)[1-f_\text{FD}(\epsilon)]\right)^{-1/3},
\label{eq:R}
\end{equation}
where $f_\text{FD}$ is the Fermi-Dirac distribution function and $g(\epsilon)$ the DOS distribution. Eq.~(\ref{eq:R}) is commonly employed to identify the critical hop in the infinite percolating network\cite{Mott2012,Baranovskii2018}, while it is used here as a measure of the range of all hops starting at energy $E$. In the former case, the parameter $B=2.8$ reflects the critical number of bonds on the percolating network. The total escape rate can then be determined from the differential escape rate [Eq.~(\ref{eq:vesc1})] via 
\begin{equation}
\nu_\text{esc}(E_0)=\int_{0}^{\infty} dW\, \nu_\text{esc}'(E_0,W).
\label{eq:vesc2}
\end{equation}
Moreover, the mean level at which the carrier is released from its initial state,
\begin{equation}
E_\text{esc}(E_0) = E_0+\frac{\int_{0}^{\infty} dW\, W \nu_\text{esc}'(E_0,W)}{\int_{0}^{\infty} dW\, \nu_\text{esc}'(E_0,W)},
\label{eq:Eesc}
\end{equation}
can be calculated with the help of the differential escape rate distribution.

From the differential form of the generalized Einstein relation one obtains the energy-dependent carrier mobility,
\begin{equation}
\mu(E) = \frac{q}{k_\text{B}T} \eta (1-f_\text{FD}) D(E),
\label{eq:mu}
\end{equation}
where $\eta$ is a fitting constant \cite{Ambegaokar1971} and $D(E) = \lambda(E)^2 \nu_\text{esc}(E)$ is the diffusion coefficient. Here we set $\eta$ equal to unity. The equivalence of Eq.~(\ref{eq:mu}) to the generalized Einstein equation is shown in the Supplemental Material\cite{SM}. The carrier mean hopping distance $\lambda(E) = \bar{R}[E_\text{esc}(E)]$ is determined by Eq.~(\ref{eq:R}). As a side note, we would like to stress that in principle the expectation value for the squared hopping distance $<R^2>$ should be used instead of $<R>^2$ \cite{Arkhipov2002}. However, the difference is negligible and did not affect the results here. 

With the energy-dependent differential conductivity $\sigma'(E) = q\, g(E)\, f_\text{FD}(E)\, \mu(E)$ at hand, the macroscopic observables $\sigma$, $S$ and $\kappa_\text{el}$ can be determined from Eq.~(\ref{eq:sigma})-(\ref{eq:kappael}). In contrast, assuming e.g. a delta function for $\sigma(E)\,\propto\,\delta(E-E^*)$ as typically done in Mott-type percolation models \cite{Mott2012,Zuo2019} leads to $\kappa_\text{el}=0$. 

A critical point in the analytical model described above is the use of an individual percolation condition for each hop (see Eq.~(\ref{eq:R})), while commonly the critical hop is taken to be the dominant hop in the infinite percolating network\cite{Mott2012,Baranovskii2018}. In view of this, and the further approximations used in treatment of the Coulomb trapping, we compare the analytical model to numerically exact kinetic MC simulations. Note that in the following the electronic contribution to the thermal conductivity obtained by the analytical model was divided by an empirical factor of 1.3, which provides a more accurate reproduction of the Lorenz number obtained from quasi-atomistic kMC simulations. 

\subsection{Density of States (DOS)}
The energy distribution of the localized sites through which hopping takes place is typically assumed to be Gaussian in shape,
\begin{equation}
g_i(E) = \frac{N_i}{\sqrt{2 \pi \sigma_\text{DOS}^2}}\exp\left(-\frac{(E-E_i)^2}{2\sigma_\text{DOS}^2}\right),
\label{eq:gaussDOS}
\end{equation}
where $E_i$ and $\sigma_\text{DOS}$ are the central energy and the width of the Gaussian DOS, respectively, and $N_i$ is the total site density. Ionized dopants are known to act as Coulomb traps and therefore lead to a perturbation of the DOS in the form of exponential tail states and broadening of the main DOS peak. Such modifications of the DOS shape can have an enormous impact on the thermoelectric properties, e.g. result in a qualitative change of the $S$ vs. $\sigma$ curve \cite{Abdalla2017,Boyle2019}. Arkhipov \textit{et al}., \cite{Arkhipov2005} developed an approximation for the ion-perturbed DOS of a doped semiconductor,
\begin{equation}
g(E) = A \int_{-\infty}^{0} \frac{dE_\text{C}}{E_\text{C}^4}\exp\left(\frac{A}{3E_\text{C}^3}\right)g_i(E-E_\text{C}),
\label{eq:gaussDopedDOS}
\end{equation}
where $A=\frac{4\pi q^6 N_\text{d}}{(4\pi \varepsilon_\text{0}\varepsilon_\text{r})^3}$, $N_\text{d}$ is the concentration of dopants, $E_\text{C}= -q^2/(4\pi \varepsilon_\text{0}\varepsilon_\text{r}r)$ is the Coulomb energy, and $\varepsilon_\text{r}$ is the relative dielectric constant of the semiconductor. Here, we use an extension to Eq.~(\ref{eq:gaussDopedDOS}) developed by Zuo \textit{et al}.\cite{Zuo2016,Zuo2019}, to account for energy level differences between the dopant and the semiconductor $\Delta E = E_\text{d}-E_i$, with $E_\text{d}$ being the relevant energy level of the dopant: 

\begin{subequations}
\begin{align}
g(E) &= \left(1-\frac{4\pi N_\text{d}}{3 N_\text{i}}\right) \frac{g_1(E)}{\int_{-\infty}^{0} dE\, g_1(E)} + \frac{4\pi N_\text{d}}{3 N_\text{i}}\frac{g_2(E)}{\int_{-\infty}^{0} dE\, g_2(E)}\\
g_1(E) &= A \int_{E_1}^{0}\frac{dE_\text{C}}{E_\text{C}^4} \exp\left(\frac{A}{3E_\text{C}^3}\right)g_i(E-E_\text{C})\\
g_2(E) &= A \int_{-\infty}^{E_1}\frac{dE_\text{C}}{E_\text{C}^4} \exp\left(\frac{A}{3E_\text{C}^3}\right)g_i(E-\Delta E -E_\text{C})
\end{align}
\label{eq:Zuo}
\end{subequations}
where $E_1=E_\text{C}(N_\text{i}^{-1/3})$ is the Coulomb energy one lattice constant away from the ionized dopant. Here we use $\Delta E = 0$, so that Eq.~(\ref{eq:gaussDopedDOS}) describes the DOS. 

For the simulations below, a standard parameter set with an attempt to hop frequency $\nu_0=\SI{E14}{s^{-1}}$, a Gaussian disorder of $\sigma_\text{DOS}=2k_\text{B}T$, an intersite distance of $a_\text{NN}=\SI{1.8}{nm}$, a localization length of $\alpha=\SI{0.36}{nm} (= a_\text{NN}/5)$, a temperature $T=\SI{300}{K}$ and dielectric constant of $\varepsilon_r=3.6$ is used unless indicated otherwise.

\section{Results}
\begin{figure}[h!]
\centering
\includegraphics{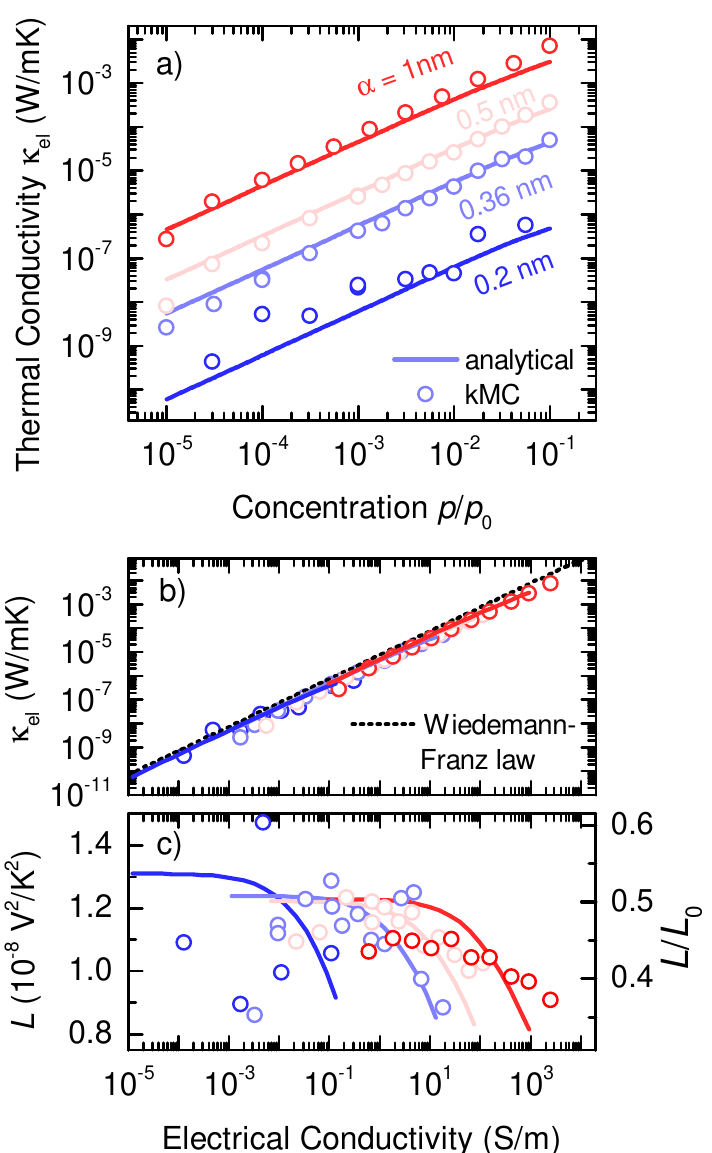}
\caption{Electronic contribution of the thermal conductivity for different localization lengths $\alpha$, calculated from the analytical hopping model (solid lines, Eq.~(\ref{eq:kappa0})--(\ref{eq:kappael})) and kMC model (symbols). Dependency on a) carrier concentration and b) electrical conductivity. The dotted black line in b) represents the result of the Wiedemann-Franz law with $L=L_0$. c) Lorenz number $L=\kappa_\text{el}/(\sigma \cdot T)$ as a function of the electrical conductivity.}
\label{fig:figure01}
\end{figure}
In Fig.~\ref{fig:figure01}a), the electronic contribution of the thermal conductivity calculated from the analytical model for various values of the localization length $\alpha$ is compared to the results from the kMC simulation. We find that both the absolute value as well as the concentration dependence of the kMC data are accurately reproduced by the analytical model. For increasing carrier concentration, the thermal conductivity $\kappa_\text{el}$ increases, in contrast to the trends observed by Lu \textit{et al}. \cite{Lu2016} but consistent with the general trends which could be expected from the Wiedemann-Franz law and are observed in experiments \cite{Liu2015,Weathers2015}. 
As shown in Fig.~\ref{fig:figure01}b), changing the localization length to larger values increases $\kappa_\text{el}$ and $\sigma$ simultaneously, as it implies stronger wavefunction overlap between adjacent sites and is therefore favorable for the hopping process. However, the scaling is not entirely linear, which results in a conductivity, and with this also carrier concentration dependent Lorenz factor, as shown in Fig.~\ref{fig:figure01}c).

\begin{figure}
\centering
\includegraphics{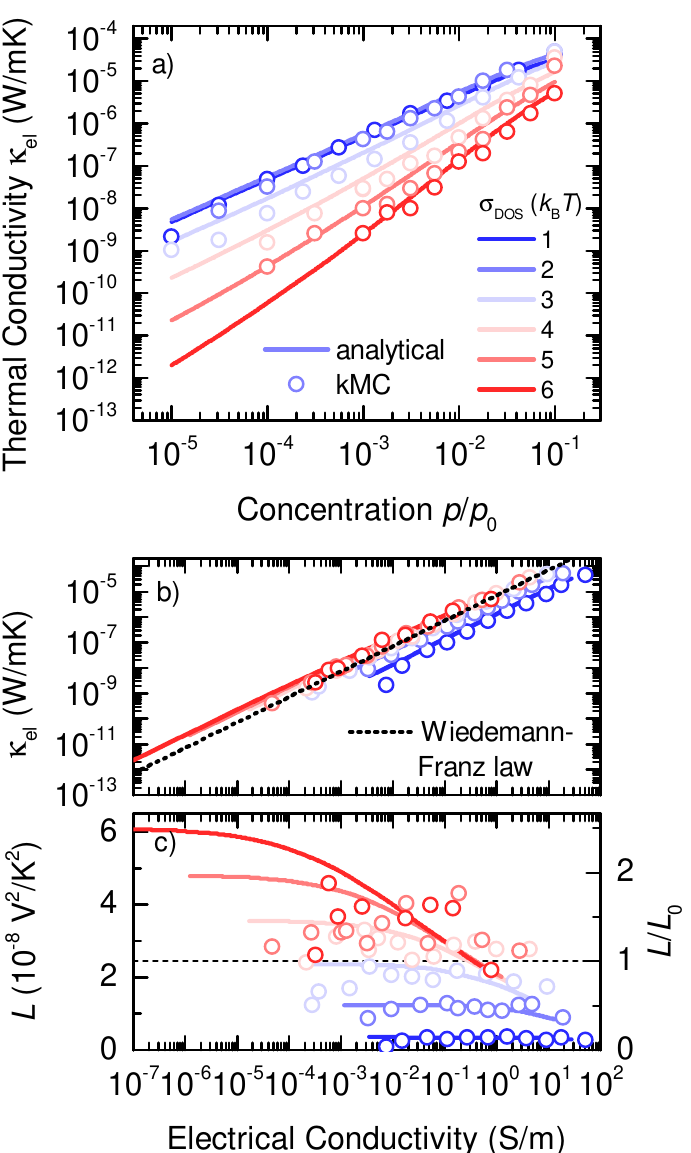}%
\caption{Electronic contribution of the thermal conductivity from the analytical model (solid lines) and kMC simulations (symbols) for different energetic disorder $\sigma_\text{DOS}$. Dependency on a) carrier concentration and b) electrical conductivity. The dotted black line in b) represents the result of the Wiedemann-Franz law with $L=L_0$. c) Lorenz number $L$ as a function of the electrical conductivity. The dashed line indicates the Sommerfeld value for a free electron gas $L_0 = \frac{\pi^2}{3}\left(\frac{k_\text{B}}{q}\right)^2$.}
\label{fig:figure02}
\end{figure}
In organic semiconductors, the disorder of the DOS is known to have a significant influence on the electrical conductivity \cite{Zuo2016} and it is therefore likely that it also impacts the thermal conductivity $\kappa_\text{el}$. Fig.~\ref{fig:figure02}a) shows that the absolute value of $\kappa_\text{el}$ as well as its dependence on the carrier concentration is strongly influenced by the disorder parameter $\sigma_\text{DOS}$. Moreover, the latter affects the slope of $\kappa_\text{el}$ vs. $\sigma$, leading to a Lorenz factor that strongly depends on the disorder. As shown in Fig.~\ref{fig:figure02}c), weakly disordered systems and high carrier concentrations would allow a Lorenz factor of that is substantially below the Sommerfeld value. This is a promising result for the development of organic TE materials, as the condition $L \ll L_0$ occurs in the high-conductivity part of the parameter space that is most relevant for TE generators. 

\begin{figure}
\centering
\includegraphics[width=0.5\textwidth]{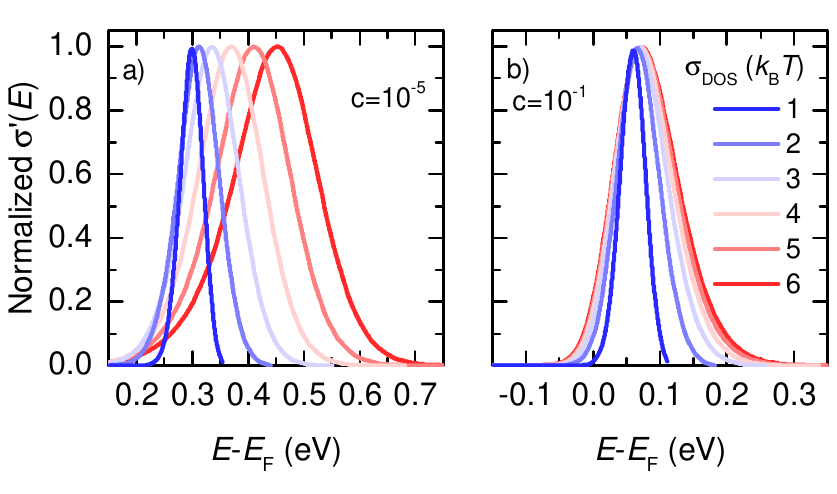}%
\caption{a) Normalized energy dependent differential conductivity calculated from the analytical model at a carrier concentration of (a) $c=10^{-5}$ and (b) $c=10^{-1}$ for different disorder parameter.}
\label{fig:figure03}
\end{figure}
The reduction in the Lorenz factor results from two variations in the energy-dependent differential conductivity $\sigma'(E)$ (see Fig.~\ref{fig:figure03}): First, a shift of the maximum of $\sigma'(E)$ to a smaller magnitude of $(E-E_F)$ with decreasing disorder, which is particularly pronounced at low carrier concentrations and affects mainly $\kappa_\text{el}$. Second, a narrowing of $\sigma'(E)$, mainly correlated to a decreasing disorder $\sigma_\text{DOS}$, which influences both $\kappa_\text{el}$ and $\sigma$ but also decreases the Lorenz factor. However, the shift of $\sigma'(E)$ has a significantly larger impact on the Lorenz factor than the broadening. The observed trend of a decreasing Lorenz factor with increasing energetic ordering is in line with findings from the inorganic community where it is known that the Lorenz factor can be lowered by reducing the bandwidth of the charge carrier dispersion \cite{Mahan1996,Jeong2012}. 
Our findings resulting from the ansatz for $\kappa_\text{el}$ are in good agreement with the few existing experimental studies. The data of Weathers \textit{et al.}\cite{Weathers2015} for PEDOT:Tosylate films lead to $L/L_0=2.5$, while PEDOT:PSS processed from DMSO shows $L/L_0\simeq 1$\cite{Liu2015} and ethylene glycol treated PEDOT:PSS $L/L_0\ll 1$\cite{Kim2013}, which is in the range of the ratio $L/L_0$ determined here. Furthermore, our results are in-line with the results shown in Ref.~\cite{Upadhyaya2019}, where the effect of disorder on the Lorenz factor in organic semiconductors is studied and with findings from the inorganic community regarding the trend of the Lorenz factor with the bandwidth of the charge carrier dispersion. Therefore, we suggest that Eqs.~(\ref{eq:kappa0}) and (\ref{eq:kappael}) provide a reasonable approximation of the electronic part of the thermal conductivity also in organic semiconductors.
\begin{figure}
\centering
\includegraphics{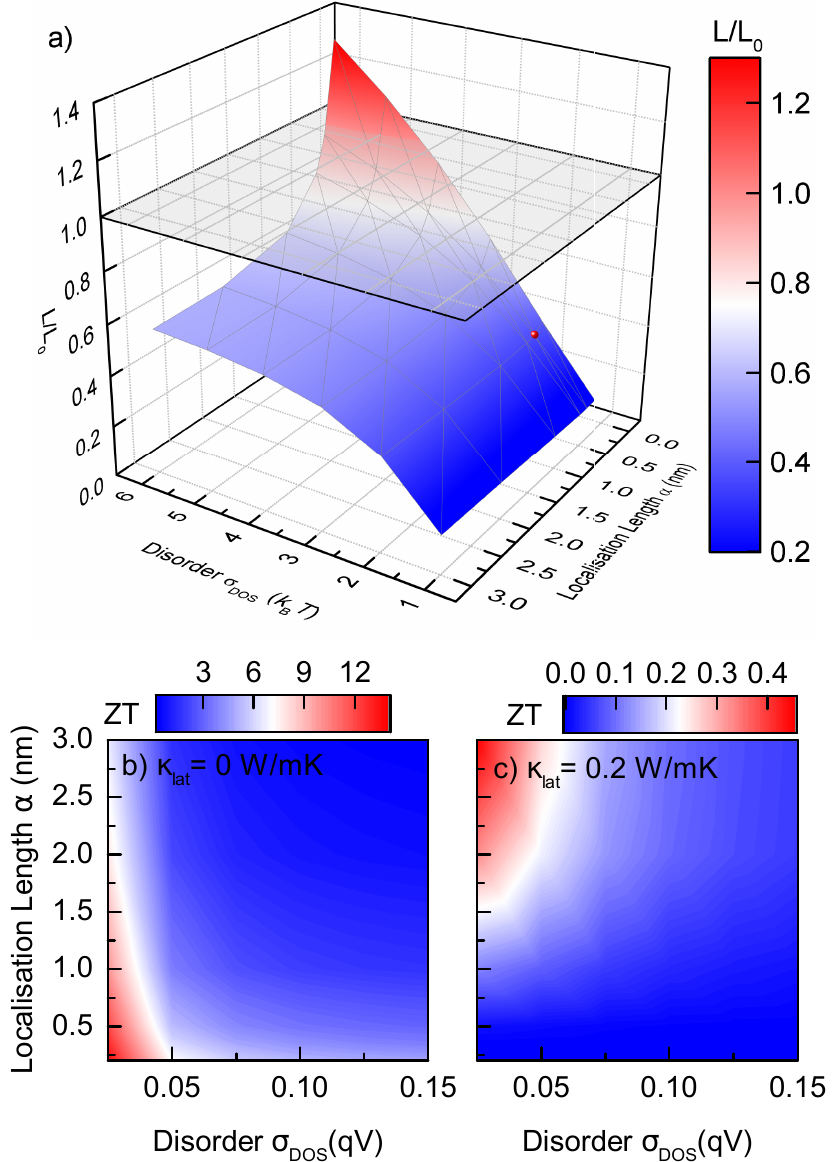}
\caption{a) Lorenz factor $L$ in units of $L_0$ and  b), c) figure of merit $ZT$ as a function of the energetic disorder $\sigma_\text{DOS}$ and the localization length $\alpha$ for a carrier concentration of 0.1. The red dot in a) represents the standard parameter set used within this study, while the plane illustrates $L=L_0$.  For b) and c) a lattice thermal conductivity of b) $\kappa_\text{lat} = \SI{0.2}{W/mK}$ and c) $\kappa_\text{lat} = \SI{0}{W/mK}$ was used.}
\label{fig:figure04}
\end{figure}

Having shown in the previous section that the analytical model can well describe the results of the kMC simulation for a purely Gaussian DOS, we used the former model to study the Lorenz factor and the figure of merit $ZT$ as a function of the energetic disorder and the localization length. We focused here on the behavior of $ZT$ at $T=\SI{300}{K}$, as the performance around room temperature is probably the most relevant case for organic TE materials in contrast to the high-temperature regime in which traditional inorganic crystalline TE materials are applied\cite{Russ2016,Campoy-Quiles2019}. Moreover, polaronic effects in terms of Marcus theory where not explicitly taken into account. Nevertheless, we have thus far been able to very well describe experimental data without explicitly accounting for polaronic effects, using the simpler Miller-Abrahams rates\cite{Abdalla2017}. Similar conclusions regarding the minor differences between the two rates were also drawn by Cottaar \textit{et al.}\cite{Cottaar2012} and Mendels and Tessler \cite{Mendels2014}. Fig.~\ref{fig:figure04}a) shows that the localization length has no significant influence on $L$ for low energetic disorder, but gains importance for $\sigma_\text{DOS}>3k_\text{B}T$. Meanwhile, the energetic disorder impacts the Lorenz factor in all parameter combinations considered within this study.

Independent on the exact value of the lattice contribution to the thermal conductivity $\kappa_\text{lat}$, a low energetic disorder is decisive for an optimized figure of merit $ZT$, as depicted in Fig.~\ref{fig:figure04}b) and c). However, whether a low or high localization length is favorable depends on $\kappa_\text{lat}$, as the absolute value of $\kappa_\text{el}$ found with this parameter set is approximately two orders of magnitude lower than $\kappa_\text{lat}=\SI{0.2}{W/mK}$, which is typically assumed for organic TE materials to date \cite{Kang2017}. 
In the limit $\kappa_\text{lat}\gg \kappa_\text{el}$, $ZT$ is proportional to the power factor $PF = S^2\sigma$ (see Supplemental Material~\cite{SM}, Fig.~S2a) and maximized for large localization length, as this leads to (exponentially) higher $\sigma$, while $S$ is hardly affected (see Supplemental Material~\cite{SM}, Fig.~S1).
In contrast, for $\kappa_\text{lat}\ll \kappa_\text{el}$, $ZT$ is given by by $PF/\kappa_\text{el} T = S^2/L$. 
Thus, in this limit the thermopower is decisive as the decrease in $S^2$ with increasing $\alpha$ is much greater than the one in $L$ (see Supplemental Material~\cite{SM}, Fig.~S2b). Note, that the lattice thermal conductivity $\kappa_\text{lat}$ and the localization length $\alpha$ could be coupled via the strength of the inter- and intra-molecular bond but are here considered as independent parameters. An important implication of Figs.~\ref{fig:figure04}b) and c) is that it appears unlikely that organic semiconductors will reach $ZT$ values approaching unity while maintaining a lattice thermal conductivity of $\sim \SI{0.2}{W/mK}$. Although the latter is already a rather low value, achieving application-relevant $ZT$ will require even lower values.

\begin{figure}
\centering
\includegraphics{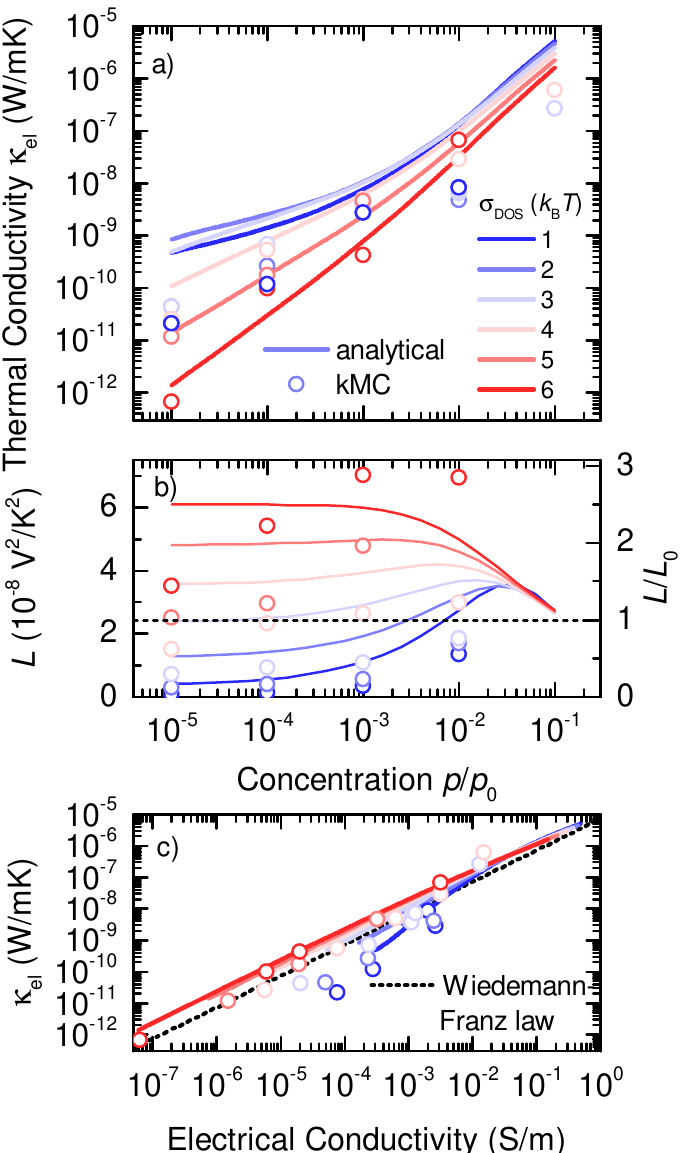}%
\caption{Electronic contribution of the thermal conductivity for doped organic semiconductors from the analytical model (solid lines) and kMC simulations (symbols) for different initial energetic disorder $\sigma_\text{DOS}$. Dependency on a) doping concentration and c) electrical conductivity.  b) Lorenz number $L$ as a function of the doping concentration.}
\label{fig:figure05}
\end{figure}

Chemical or electrochemical doping of organic semiconductors is an effective way to increase the electrical conductivity by increasing the charge carrier concentration and is increasingly used in applications. As discussed above, adding ionized dopants to the intrinsic semiconductor will change the shape of the DOS due to long-range Coulombic interactions. Figure~\ref{fig:figure05} shows the electronic thermal conductivity $\kappa_\text{el}$ calculated for a broadened DOS (see Eq.~(\ref{eq:Zuo})) for various values of the energetic disorder $\sigma_\text{DOS}$. While in case of the electrical conductivity the analytical model still describes the results from kMC simulations reasonably well (see Supplemental Material~\cite{SM}, Fig.~S3a), this is not the case for the thermal conductivity, as shown in Fig.~\ref{fig:figure05}a) and the thermopower (see Supplemental Material~\cite{SM}, Fig.~S3b). However, the general trends, i.e. a Lorenz number of $L<L_0$ for $\sigma_\text{DOS}<3k_\text{B}T$ (see Fig.~\ref{fig:figure05}b) and c)) are reproduced by the analytical model, which results from a cancellation of errors in $\kappa_\text{el}$ and $\sigma$.
Note that a first reaction scheme was used for the kMC simulations in Fig.~\ref{fig:figure05}, meaning that rates and energies were only recalculated for the moving particle, which significantly reduces calculation time. This approximation is accurate up till relative doping levels of $\simeq 10^{-2}$, as shown by Zuo \textit{et al}. \cite{Zuo2019}. 

\begin{figure}
\centering
\includegraphics{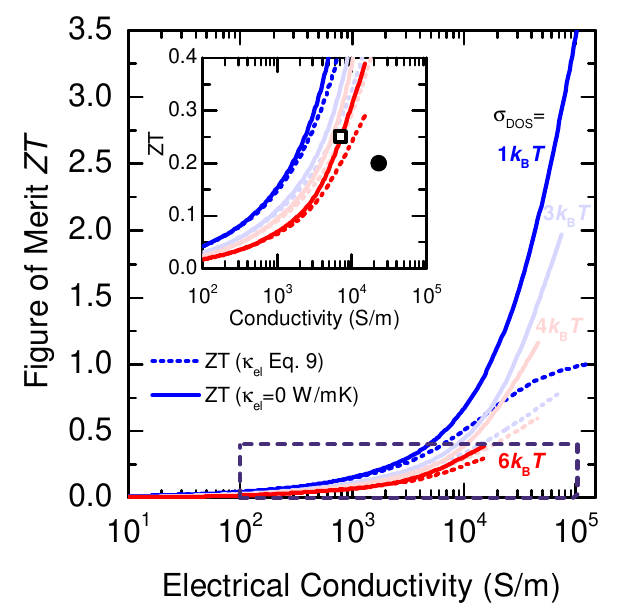}%
\caption{Figure of merit as a function of the electrical conductivity for different energetic disorder values and with $\kappa_\text{el}$ determined by Eq.~(\ref{eq:kappael}) (dotted lines) and $\kappa_\text{el}=0$ (solid lines) to illustrate the influence of the electronic contribution to the thermal conductivity. Charge carriers are introduced by doping and parameters are set according to the standard parameter set but with $\nu_0=\SI{3E16}{s^{-1}}$ and $\kappa_\text{lat}=\SI{0.2}{W/mK}$. The inset shows a zoom-in to the simulated data and experimental data for (n-type) poly(nickel-ethylenetetrathiolate) (black circle, ref.~\cite{Sun2016}) and for (p-type) PEDOT-Tos (black square, ref.~\cite{Bubnova2011}).}
\label{fig:figure06}
\end{figure}

Extrapolating the electrical conductivity in Fig.~\ref{fig:figure05} to $\sigma=\SI{100}{S/cm}\,(=\SI{E4}{S/m}$) leads to $\kappa_\text{el}\approx 0.05-\SI{0.1}{W/mK}$ which is in the same range as the experimental value of $\kappa=0.3\pm \SI{0.1}{W/mK}$ observed by Bubnova \textit{et al.}\cite{Bubnova2011} and Kim \textit{et al.} \cite{Kim2013}, when taking into account $\kappa_\text{lat}\approx\SI{0.2}{W/mK}$. However, comparing the magnitude of the calculated $\kappa_\text{el}$ in Figs.~\ref{fig:figure01}, \ref{fig:figure02} and \ref{fig:figure05} to the typically used (estimated) $\kappa_\text{lat}=\SI{0.2}{W/mK}$ suggests that the lattice contribution to the thermal conductivity will dominate in most experimental systems that are currently under investigation. This is further illustrated in Fig.~\ref{fig:figure06} that shows the figure of merit $ZT$ from the analytical model with $\kappa_\text{el}=\SI{0}{W/mK}$ (solid lines) and $\kappa_\text{el}$ determined by Eq.~(\ref{eq:kappael}) (dotted lines). Also shown in the inset of Fig.~\ref{fig:figure06} are current room temperature records for p-type materials (PEDOT:Tos with $ZT\sim 0.25$ (black square), \cite{Bubnova2011}) and n-type materials (poly(nickel-ethylenetetrathiolate) with $ZT\sim 0.2$ (black circle) \cite{Sun2016}).

Within the part of the parameter space up till the current record values, the impact of $\kappa_\text{el}$ on $ZT$ is small, as seen by the small difference between the dashed and solid lines in the inset of Fig.~\ref{fig:figure06}. However, for commercial applications the conductivity of these materials needs to be improved further, which would lead to a significant contribution of $\kappa_\text{el}$ to $ZT$. At the same time, the lattice contribution $\kappa_\text{lat}$ becomes a limiting factor (see Supplemental Material~\cite{SM}, Fig.~S4) even for the most optimistic case of $\sigma_\text{DOS}=1k_\text{B}T$ (blue dashed line), in line with the discussion at Fig.~\ref{fig:figure04} above. Hence, in order to reach and surpass unity $ZT$, it will likely be needed to suppress the lattice contribution to the thermal conductivity of organic semiconductors. This situation is extremely reminiscent of that in high-performance inorganic semiconductor materials \cite{Hu2014,Shakouri2011,Kanatzidis2017}. 

\section{Conclusion}
The electrical contribution to the thermal conductivity of disordered organic semiconductors has been systematically analyzed by means of analytical modeling and numerical kinetic Monte Carlo simulations. We show that optimizing organic thermoelectrics with respect to thermal properties requires to consider their energetic disorder as well as the relevant length scales of the system, i.e. the localization length relative to the typical inter-site distance. Large derivations from the Wiedemann-Franz law can be observed that can be characterized by an effective Lorenz number $L$ that can be larger as well as substantially smaller than the Sommerfeld value $L_0$. In consequence, our results are in contradiction with the universal Lorenz number found for materials with band transport but can give an explanation for the large range of Lorenz numbers reported from experimental studies. Minimizing energetic disorder can be a viable strategy to reduce the Lorenz factor and obtain higher thermoelectric performance. Moreover, we show that for most cases reported to date $\kappa_\text{el} <\kappa_\text{lat}$ but the electronic contribution to the thermal conductivity will unavoidably dominate the thermal conductivity for higher electrical conductivities that are needed to reach application-relevant figures of merit. At the same time, reaching beyond $ZT = 1$ will likely require suppressing the lattice thermal conductivity to values below $\kappa_\text{lat}=\SI{0.2}{W/mK}$, a situation that is conceptually similar to that for inorganic thermoelectrics.

\begin{acknowledgement}
This project has received funding from the European Union’s Horizon 2020 research and innovation program under the Marie Sk\l{}odowska-Curie grant agreement No 799477 — HyThermEl.
\end{acknowledgement}

\bibliography{references}

\end{document}